\documentclass[nofootinbib,showpacs,preprintnumbers,amsmath,amssymb,floatfix]
{revtex4}
\usepackage{epsf}

\begin{document}

%\begin{center}

\title{Parametrization of the QCD coupling in Hard and Regge processes}

\vspace*{0.3 cm}

\author{B.I.~Ermolaev}
\affiliation{Ioffe Physico-Technical Institute,194021
 St.Petersburg, Russia}
\author{S.I.~Troyan}
\affiliation{St.Petersburg Institute of Nuclear Physics, 188300
Gatchina, Russia}

\begin{abstract}
We examine the parametrization of the QCD coupling in the Bethe-Salpeter
equations for the hard and Regge processes and determine the argument
of $\alpha_s$ of the factorized gluon.
Our analysis shows that for the hard processes
$\alpha_s = \alpha_s(k^2_{\perp}/(1- \beta))$ where $k^2_{\perp}$
and $\beta$ are
the longitudinal and transverse moment of the soft parton.
On the other hand, in the Regge processes
$\alpha_s = \alpha_s(k^2_{\perp}/\beta)$. We have also shown
that the well-known parametrization $\alpha_s =
\alpha_s(k^2_{\perp})$ in the DGLAP equations
stands only if the lowest
integration limit $\mu^2$ over $k^2_{\perp}$ (the starting point
of the $Q^2$ -evolution) obeys the relation $\mu \gg \Lambda_{QCD} \exp {(\pi/2)}$,
otherwise the coupling should be replaced by the more complicated expression.
\end{abstract}


\maketitle

\section{Introduction}
Total resummations of radiative corrections in QCD are often performed with
composing and solving evolution equations. Quite often such equations
are of the Bethe-Salpeter type, with one gluon (or one ladder rung) being factorized.
In particular, both the DGLAP and BFKL equations are of that type.
As is well-known, the argument
of $\alpha_s$ in the DGLAP equations is $Q^2$. It follows from the dependence
\begin{equation}\label{adglap}
\alpha_s = \alpha_s(k^2_{\perp})
\end{equation}
 in each rung of the ladder Feynman graphs. Such a
dependence originally was used in analogy to the parametrization of $\alpha_s$ in
the hard kinematics according to the results of Refs.~\cite{ddt,av}. Later,
the proof of the parametrization (\ref{adglap}) in the DGLAP context was suggested
in Ref.~\cite{ds}. However, that proof was not done accurate enough.
In our recent paper Ref.~\cite{etalfa} we have revised Ref.~\cite{ds} and showed
that $\alpha_s$ in the Bethe-Salpeter equations, including
DGLAP, is basically replaced by the effective coupling $\alpha_s^{eff}$
given by much more complicated expression than Eq.~(\ref{adglap}):
\begin{equation}\label{aeff}
\alpha_s^{eff} = \alpha_s(\mu^2) + \frac{1}{\pi b}
\arctan \Big(\frac{\pi [\ln (k^2_{\perp}/\beta \Lambda^2) -
\ln (\mu^2/\Lambda^2)]}
{\pi^2 + \ln (k^2_{\perp}/\beta \Lambda^2)\ln (\mu^2/\Lambda^2)}\Big)
\end{equation}
where the Sudakov variable $\beta$ is the fraction of the longitudinal
of the ladder parton, $\Lambda = \Lambda_{QCD}$ and $b = [11 N - 2 n_f]/(12 \pi)$.
However when $\mu$, the starting point of the $Q^2$- evolution is chosen large
enough, namely when
\begin{equation}\label{api}
\mu \gg \Lambda_{QCD} e^{\pi/2},
\end{equation}
Eq.~(\ref{aeff}) can can be simplified down to
\begin{equation}\label{aeffpi}
\alpha_s^{eff} \approx \alpha_s(k^2_{\perp}/\beta).
\end{equation}
When Eq.~(\ref{aeff}) is applied to the DIS structure functions at $x \sim 1$
and Eq.~(\ref{api}) is also fulfilled, Eqs.~(\ref{aeff},\ref{aeffpi}) can be approximated by
the DGLAP expression of Eq.~(\ref{adglap}).

\section{Parametrization of $\alpha_s$ for QCD processes in the hard kinematics}

\begin{figure}[!thb]
\begin{center}
\begin{picture}(125,150)
\put(0,0){ \epsfbox{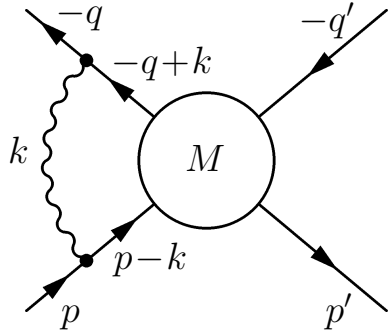} }
\end{picture}
\end{center}
\caption{The right-hand side of Eq.~(\ref{mssud})}
\label{alfafig1}
\end{figure}

Let us consider the contribution $M_t$ of the Feynman graph
depicted in Fig.~\ref{alfafig1}.
The cases with $u$ and $s$ -channel gluons factorized can be
considered quite similarly.
 The solid lines in Fig.~\ref{alfafig1} denote quarks, though
the generalization to the case of gluons is obvious. Through the
paper we will assume that the lower particles, with momenta
$p_1,~p'_1$, have small virtualities $\sim \mu^2$ whereas
virtualities of the upper partons, with momenta $q,~q'$ are large:
$- q^2 \sim -q'^2\sim Q^2 \gg \mu^2$. Applying the Feynman rules
to Fig.~\ref{alfafig1} we obtain:
\begin{equation}\label{mssud}
M_t = \imath \frac{s^2}{4 \pi^2} \int d \alpha d \beta d k^2_{\perp}
\frac{M \big(s, Q^2,s\alpha, s \beta, k^2_{\perp}\big)}
{[s \beta - Q^2 - s \alpha\beta - k^2_{\perp} + \imath \epsilon]
[s \alpha- s \alpha\beta - k^2_{\perp} + \imath \epsilon]}
\frac{\alpha_s (- s \alpha\beta - k^2_{\perp})}{(s \alpha\beta + k^2_{\perp} - \imath \epsilon)}.
\end{equation}
We have used the standard Sudakov variables $k = - \alpha (q +xp)
+ \beta p + k_{\perp}$ and we have dropped the color factors as
unessential for our analysis. $M$ corresponds to the blob in
Fig.~\ref{alfafig1}. In Eq.~(\ref{mssud}) we have also neglected
the virtuality $p^2 = \mu^2$ of the initial parton and denoted $s
= 2pq,~ x = Q^2/2pq,~Q^2 = - q^2$. Amplitude $M$ is unknown, so it
is impossible to perform the integration over any of the variables
in Eq.~(\ref{mssud}). However, if we assume the leading
logarithmic (LL) accuracy, we can use the QCD -generalization of
the bremsstrahlung Gribov theorem. According to it, $M$ does not
depend on $\alpha$ and $\beta$. Integrating over $\alpha$ in
Eq.~(\ref{mssud}) is conventionally performed with closing the
integration contour down and taking the residue at $s\alpha = (+
k^2_{\perp} - \imath \epsilon)/(1- \beta)$. It converts
Eq.~(\ref{mssud}) into
\begin{equation}\label{ahard}
M_t = - \frac{1}{2 \pi} \int_{\mu^2}^s \frac{d k^2_{\perp}}{k^2_{\perp}}
\int_{\beta_0}^1 d \beta\frac{(1- \beta)}{\beta} M(s,Q^2, k^2_{\perp})
\alpha_s \Big(- \frac{k^2_{\perp}}{(1-\beta)} \Big)
\end{equation}
where $\beta_0 = x + k^2_{\perp}/s$. Obviously, $\beta_0 \approx x = Q^2/2pq$
when $x \sim 1$ and the upper limit $s$ of the integration over
$k^2_{\perp}$ can be changed for $Q^2$.
The minus
sign of the $\alpha_s$ -argument in Eq.~(\ref{ahard}) indicates explicitly that
the argument is space-like and for the space-like argument $\alpha_s$ is given
by the well-known expression:
\begin{equation}\label{alphahard}
\alpha_s \Big(- \frac{k^2_{\perp}}{(1-\beta)} \Big) =
\frac{1}{b \ln \Big(k^2_{\perp}/\big((1-\beta)\Lambda^2\big)\Big)}.
\end{equation}
With the LL
accuracy, $k^2_{\perp}/(1 - \beta) \approx k^2_{\perp}$ and therefore in Eq.~(\ref{ahard})
$\alpha_s \approx \alpha_s (k^2_{\perp})$. The minus sign of the argument of
$\alpha_s$ is traditionally
dropped, which drives us back to the standard expression of Eq.~(\ref{adglap}).

\section{Parametrization of $\alpha_s$ in the Bethe-Salpeter equations}

\begin{figure}[!thb]
\begin{center}
\begin{picture}(125,150)
\put(0,0){ \epsfbox{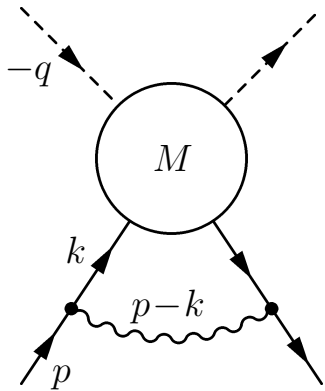} }
\end{picture}
\end{center}
\caption{The integral contribution in Eq.~(\ref{aeq})}
\label{alfafig2}
\end{figure}

In this section we study the parametrization of $\alpha_s$ in the Bethe-
Salpeter equation for the forward scattering amplitude $A$ .
Let us assume that $A$ obeys the following
Bethe-Salpeter equation:

\begin{eqnarray}\label{aeq}
A
%=A_0 + \frac{\imath}{4 \pi^2}\int d \alpha d \beta d k^2_{\perp}
%M ((q+k)^2, Q^2, (s\alpha\beta + k^2_{\perp}))
%\frac{sk^2_{\perp}}{(-s\alpha\beta - k^2_{\perp} + \imath \epsilon)^2}
%\frac{\alpha_s (s\alpha(1-\beta) - k^2_{\perp})}{[s\alpha(1-\beta) - k^2_{\perp} +
%\imath \epsilon]} \\ \nonumber
 = A_0 + \frac{\imath}{4 \pi^2}\int d k^2_{\perp} d \beta d m^2
 M (s \beta, Q^2,~ (m^2\beta + k^2_{\perp}))
\frac{(1- \beta)k^2_{\perp}}{(m^2\beta + k^2_{\perp} - \imath \epsilon)^2}
\frac{\alpha_s (m^2)}{[m^2 + \imath \epsilon]}.
\end{eqnarray}
The second term in the rhs of Eq.~(\ref{aeq}) is depicted in
Fig.~\ref{alfafig2}.
We have used the standard Sudakov variables: $k = - \alpha (q +xp)
+ \beta p + k_{\perp}$. Following Ref.~\cite{ds}, we have replaced
the Sudakov variable $\alpha = 2pk/2pq$ by the new variable $m^2 =
(p-k)^2$. $M$ in Eq.~(\ref{aeq}) denotes the upper blob in
Fig.~\ref{alfafig2}. It includes both the off-shell amplitude $A$
and a kernel.
 Now we just
notice that Eq.~(\ref{aeq}) can be solved only after $M$ has been known. $A_0$
stands for an inhomogeneous term. We focus on integrating  over $\alpha$ in
Eq.~(\ref{aeq}) and introduce

\begin{equation}\label{i}
I = \int_{- \infty}^{\infty} d m^2 M (s \beta, Q^2,~ (m^2\beta + k^2_{\perp}))
\frac{(1- \beta)k^2_{\perp}}{(m^2\beta + k^2_{\perp} - \imath \epsilon)^2}
\frac{\alpha_s (m^2)}{[m^2 + \imath \epsilon]}.
\end{equation}

\begin{figure}[!thb]
\begin{center}
\begin{picture}(125,150)
\put(0,0){ \epsfbox{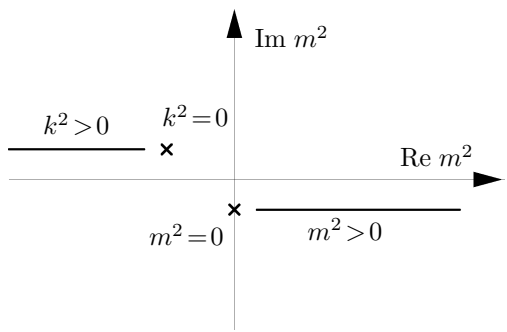} }
\end{picture}
\end{center}
\caption{Singularities of $I$ given by Eqs.~(\ref{mpole} -
\ref{acut})} \label{alfafig3}
\end{figure}

The integrand of Eq.~(\ref{i}) has the singularities in $m^2$.
First, there are two poles from the propagators:
\begin{equation}\label{mpole}
m^2  = -k^2_{\perp}/\beta + \imath \epsilon
\end{equation}
and
\begin{equation}\label{apole}
m^2 = 0 - \imath \epsilon.
\end{equation}
Second, there are two cuts. The first cut is originated by the
$k^2$ -dependence of $M$. In particular, it can be the logarithmic
dependence. The cut begins at
\begin{equation}\label{mcut}
m^2  = -k^2_{\perp}/\beta + \imath \epsilon
\end{equation}
and goes to the left. The second cut is related to $\alpha_s$. It begins at
\begin{equation}\label{acut}
m^2 = 0 - \imath \epsilon
\end{equation}
and goes to the right. The singularities (\ref{mpole}-\ref{acut})
are depicted in Fig.~\ref{alfafig3}.
The integration over $m^2$ in Eq.~(\ref{i}) runs along the $\Re
m^2$ -axis from $ - \infty$ to $ \infty$, so the integral can be
calculated with choosing an appropriate closed integration contour
$C$ and taking residues. The contour $C$ should include the line
$- \infty < m^2 < \infty$ and a semi-circle $C_R$ with radius $R$.
The contour $C_R$ may be situated either in the upper or in the
right semi-plane of the $m^2$ -plane. However, if we choose $C_R$
to be in the upper semi-plane, we should deal with the cut
(\ref{mcut}) of an unknown amplitude $M$, which is impossible
without making assumptions about $M$.
%Such a closing of the contour was chosen
%in Ref.~\cite{ds} but the contribution of the cut (\ref{mcut}) was not
%taken into account because there was made the assumption that
%$M(s \beta, Q^2, k^2)  \approx M(s \beta, Q^2, k^2_{\perp})$,
%i.e. that $k^2_{\perp} \gg |m^2\beta|$. This assumption made possible
% to calculate the integral $I$ with taking the
%residue at the pole Eq.~(\ref{mcut}) where $k^2_{\perp} = |m^2\beta|$.
%This obvious contradiction between the assumption
%  and (\ref{mpole})  makes the method of Ref.~~\cite{ds} inconsistent.
Alternatively,
choosing the contour $C_R$ in the lower semi-plane involves analysis of
the cut (\ref{acut}) of  $\alpha_s$ and $\alpha_s$ is known.
%given by Eq.~(\ref{alphas}).
By this reason, we choose the latter option for $C_R$.
%So, as shown in Fig.~4,
%the closed contour $C$ includes
%the line $- \infty < m^2 < \infty$, the lower semi-circle  $C_R$  and
the contour $C_{cut}$ which runs along both sides of the cut (\ref{acut}).
According to the Cauchy theorem,
\begin{eqnarray}\label{icont}
I_C \equiv \int_C d m^2  M (s \beta, Q^2, k^2))
\frac{(1-\beta)k^2_{\perp}}{(m^2\beta + k^2_{\perp} - \imath \epsilon)^2}
\frac{\alpha_s (m^2)}{[m^2 +
 \imath \epsilon]} =
% \\ \nonumber
 -2 \pi \imath \frac{(1-\beta)}{k^2_{\perp}}
 M (s \beta, Q^2, -k^2_{\perp}/(1-\beta)) \alpha_s (\mu^2).
\end{eqnarray}
The rhs of Eq.~(\ref{icont}) is
the residue at the pole (\ref{apole}) and  $\mu$ is introduced to regulate
the IR singularity for $\alpha_s$. It should be chosen as large as $\mu >> \Lambda$
to guarantee applicability of the perturbative expression
for $\alpha_s$. When the initial
partons are quarks, $\mu$ should also obey $\mu \gg$ the quark mass.
% The minus sign in the rhs of Eq.~(\ref{icont}) appears because
%of the clock-wise integration direction of the integration.
Obviously,
\begin{equation}\label{iccut}
I_C = I + I_{cut} + I_R
\end{equation}
where $I$ is defined in Eq.~(\ref{i}),
$I_R$ stands for the integration over the lower semi-circle and $I_{cut}$  refers
to the integration along the cut (\ref{acut}). $I_R$ can be dropped
because $I_R \to 0$
when $R \to \infty$. Now we specify $I_{cut}$:
\begin{equation}\label{icut}
I_{cut} = -2 \imath \int_{\mu^2}^{\infty} d m^2 M (s \beta, Q^2,~ (m^2\beta + k^2_{\perp}))
\frac{(1 - \beta)k^2_{\perp}}{(m^2\beta + k^2_{\perp} - \imath \epsilon)^2}
\frac{\Im \alpha_s (m^2)}{m^2}.
\end{equation}
The integration in Eq.~(\ref{icut}) cannot be done precisely because it involves the unknown
amplitude $M$ depending on $m^2$.
Nevertheless, it is possible to estimate $I_{cut}$. Indeed, the
$m^2$- dependence of $M$ in Eq.~(\ref{icut}) can be neglected in the region
$m^2 \ll k^2_{\perp}/\beta$.
Doing so, we obtain the following estimate of $I$:
 \begin{equation}\label{icutm}
 I \approx -2 \imath \frac{(1 - \beta)}{k^2_{\perp}}M (s \beta, Q^2, k^2_{\perp})
\int_{\mu^2}^{k^2_{\perp}/\beta} d m^2 \frac{\Im \alpha_s (m^2)}{m^2}=
-\frac{2\imath \pi (1- \beta)}{k^2_{\perp}}M (s \beta, Q^2, k^2_{\perp})
\alpha_s^{eff},
%- \frac{2\imath \pi}{b} \frac{(1- \beta)}{k^2_{\perp}}M (s \beta, Q^2, k^2_{\perp})
%\int_{\mu^2}^{k^2_{\perp}/\beta} \frac{d m^2}{m^2} \frac{1}{[\ln^2(m^2/\Lambda^2) + \pi^2]}
 \end{equation}
 with $\alpha_s^{eff}$ given by Eq.~(\ref{aeffpi}).
 When $\mu$ is chosen as large that Eq.~(\ref{api}) is fulfilled,
 we can drop $\pi^2$  in Eq.~(\ref{icutm}) and
arrive at the estimate
\begin{equation}\label{bigmu}
I \approx  -\frac{2\imath \pi (1- \beta)}{k^2_{\perp}}M (s \beta, Q^2, k^2_{\perp})
\alpha_s (k^2_{\perp}/\beta).
\end{equation}
Finally, we give several estimates of $R = |\alpha_s^{eff} - \alpha_s|/\alpha_s^{eff}$,
assuming that $\Lambda \approx 0.1$GeV:
$R = 5 \%$ at $\mu^2 \approx 30$GeV$^2$; then $R = 10 \%$ at $\mu^2 \approx 2.4$GeV$^2$
and $R = 50 \%$ at $\mu^2 \approx 0.8$GeV$^2$.

\section{Acknowledgments}
The work is partly supported by the  EU grant MTKD-CT-2004-510126 in partnership
with the CERN Physics Department and Russian State Grant for Scientific School
RSGSS-5788.2006.2.

\end{document}